# Atomistic Wear Mechanisms in Diamond: Effects of Surface Orientation, Stress, and Interaction with Adsorbed Molecules

Huong T. T. Ta[1], Nam V. Tran[2,*], M. C. Righi[1,*]

[1] Department of Physics and Astronomy, University of Bologna, 40127 Bologna, Italy

[2] School of Material Science and Engineering, Nanyang Technological University, Singapore

**AUTHOR INFORMATION**

Corresponding Author

M.C. Righi – E-Mail: clelia.righi@unibo.it, Department of Physics and Astronomy, University of Bologna, 40127 Bologna, Italy.

Nam V. Tran– E-Mail: vannam.tran@ntu.edu.sg, School of Material Science and Engineering, Nanyang Technological University, Singapore.

# ABSTRACT

Despite its unrivaled hardness, diamond can be severely worn during the interaction with others, even softer materials. In this work, we calculate from first-principles the energy and forces necessary to induce the atomistic wear of diamond, and compare them for different surface orientations and passivation by oxygen, hydrogen, and water fragments. The primary mechanism of wear is identified as the detachment of carbon chains. This is particularly true for oxidized diamond and diamond interacting with silica. A very interesting result concerns the role of stress, which reveals that compressive stresses can highly favor wear, making it even energetically favorable.

# TOC GRAPHICS

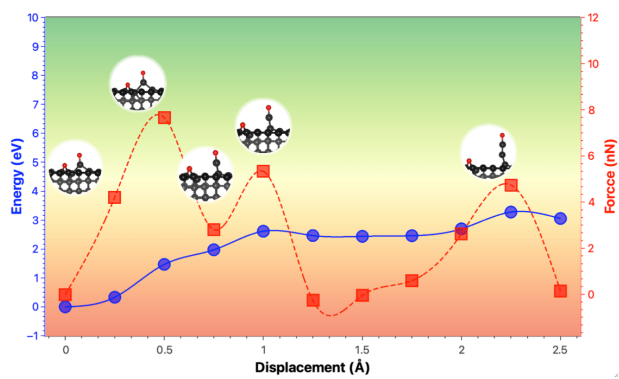

## 1. Introduction

The superior hardness of diamond provides this material exceptional wear resistance, ultra-low friction, and high durability, making it an ideal choice for various trbological applications such as cutting tools, bearings, microelectromechanical systems and coatings [1–3]. Despite the exceptional hardness, diamond is not completely inert to wear and can be worn out over a long working duration [4–7]. In these cases, either the impacts of the environment or the contributions from extreme working condition could be detrimental to diamond devices, thus causing wear, high friction, material loss, and low surface quality. To this point, understanding the wear mechanism on diamond surfaces is extremely important not only for its applications but also for the fabrication of the material.

Several studies have described the chemical modifications of diamond surfaces under different operating conditions such as high humidity, where atmospheric gases ($H_2$, $O_2$, and $H_2O$) are present. The presence of $H_2$ or $H_2O$ can create surface hydrogenation or hydroxylation, which have been shown beneficial for reducing friction and wear at the contacting interfaces [8–11]. Recently, it has been shown that the dissociation of $O_2$ is even kinetically and thermodynamically more favorable on diamond surfaces than $H_2$ and $H_2O$ [12], making surface oxidation possibly easier than hydrogenation or hydroxylation. The dissociative adsorption $O_2$ not only alters diamond surface morphology, but also modifies its electronic structure such as narrowing the band gap [13], changing the surface reactivity [14], or conductivity [15]. In particular, $O_2$ dissociation induces the breaking of surface dimer bonds, leading to the de-reconstruction on C(100) surface [16]. The presence of adsorbed O chemically modifies the diamond surfaces and facilitates the desorption in the forms of CO and $CO_2$ [17-18], which is the principle of oxidative etching and fabrication of diamond [19]. Such chemical modifications make diamond surface more or less vulnerable to external conditions, especially under harsh conditions. Thus wear is highly expected during a long duration of high loads and shear.

Great effort has been devoted to understand atomic mechanism of wear and chemical mechanical polishing process of diamond surfaces [20–23]. Several mechanisms have been proposed. $sp^3$-to-$sp^2$ transformation has been proposed to explain the experimental observation of amorphous carbon and amorphous wear particles [24]. By MD simulations, it was shown that C-C bond breaking starts at the weak C-C bonds connecting the C-C zig-zag chain to the bulk diamond when sliding against silica. In this case, the C-C bond breaking can be initiated through strong Si-C and O-C bonds where Si and O play as mechanically supported pilot atoms [25-26]. This event was not observed when sliding against silicon [19], highlighting the role of counter surfaces on diamond polishing process. By a tight-binding quantum chemical molecular dynamics simulation, Kubo et al. proposed two mechanisms of wear, i.e., atom-by-atom and sheet-by-sheet removal of diamond polishing in the presence of OH radicals acting as an oxidizing agent [23]. In $H_2O_2$ solution, ReaxFF MD showed that C can react with decomposed –H, -OH, and -O, leading to the formation of C-H, C-OH, and C-O. The C atoms can be removed in the forms of CO, $CO_2$, or carbon chain as a result of the combination of chemical and mechanical effects [21]. Another aspect is the dependency of wear on surface orientations [18, 22, 24]. While C(111) is reported as the hardest plane to be polished [24], C(001) and C(110) are more vulnerable and easier to be worn [18]. This point has been proven and clearly shown on clean surfaces. Nevertheless, the effects of oxidation and adsorption of atmospheric species on diamond polishing remain unclear. A detailed investigation of the effect of adsorbed O and other chemical groups on diamond polishing is of significance for its widespread applications in practical conditions.

In this work, we introduce a computational procedure based on first-principles calculations to characterize different atomistic mechanisms of wear and estimate the associated energy cost and forces. We applied this technique to study the effects of surface orientation (we considered the C(001), C(110), C(111) surfaces of diamond), O, H, OH adsorption, interaction with silica and surface stress (compression and expansion) on the wear mechanism. The current study will provide an understanding of the atomistic mechanisms of wear on diamond surfaces.

## 2. Computational details

In this research, Density Functional Theory (DFT) simulations were carried out using the Quantum ESPRESSO software [27]. The exchange-correlation term was described using the Generalized Gradient Approximation (GGA) as parametrized by Pedew-Burke-Ernzerhof [28]. To account for the long-range van der Waals interactions, a semi-empirical correction by Grimme (D2) was used [29 - 30]. This combination of GGA-PBE and D2 has been found to provide an optimal balance between accuracy and computational cost, as indicated by previous studies [31-32]. The convergence threshold was set at $10^{-4}$ Ry for total energy and $10^{-3}$ Ry/Bohr for ionic forces. The self-consistent electronic (SCF) loop was set to converge at $10^{-6}$ Ry. Spin polarization was included in all calculations as the presence of surface dangling bonds and dissociated molecules could result in magnetization in the system.

In this study, three distinct diamond surfaces were evaluated, namely: the C(110), the dimer-reconstructed C(001), and the Pandey-reconstructed C(111) surfaces. The simulations utilized a large orthorhombic supercell with a minimum lateral dimension of 8.74 Å in order to minimize interaction with periodic replicas. The C(110) and C(001) surfaces were modeled using a 4 × 4 in-plane supercell, while the C(111) surface was simulated using a 4 × 3 in-plane slab. The slab thickness used to model the C(110) corresponds to 7 atomic layers. Meanwhile, slabs of 8 and 10 atomic layers have been used for the C(111) and C(001) surfaces, respectively. The thickness values were selected based on previous studies [33–36], for obtaining accurate structural and energetic properties. A 20 Å vacuum region was included in the supercell to separate each slab from its periodic replicas along the [001] direction. All simulations employed a plane wave cut-off of 30 Ry and a Monkhorst-Pack (MP) k-point mesh of $2 \times 2 \times 1$. The cut-off energy, k-point mesh, and vacuum thickness were tested to ensure that the energy error was less than 3 meV/atom.

## 3. Results and discussion

We first investigate the effect of oxygen adsorption on the wear mechanism of the diamond surfaces including C(110), C(001), and C(111)-Pandey compared with the clean ones. The initial structures for the calculations are presented in **Figure S1**. The dissociative adsorptions of the oxygen molecule are considered with both Ether (oxygen atoms form C-O-C bond with the carbon atoms of diamond surfaces) and Ketone (oxygen atoms form a double bond with a single carbon atom of diamond surfaces C=O) configurations (**Figure S1)**. Previous works showed that Ketone configurations are more stable for the C(110) and C(001) at low adsorption coverage [13]. Meanwhile, the configuration is highly unfavorable for the C(111) reconstructed surface [12].We mimic the effect of wear by pulling one atom (the adsorbed O atom or C atom for the case of clean surface) along the vertical direction. In particular, for each step, the oxygen atom is displaced by 0.25 Å followed by an atomic relaxation with the position of the O atom fixed along the z direction. In **Figure 1a** the energy increase due to the displacement of the pulled atom is reported as a function of the atom-surface distance (the corresponding forces are reported in **Figure S2)**. The arrows indicate the first atom detachment from the surface. We show that very high pulling energies are required to detach the carbon atom from the clean surface. In particular, pulling energies of 2.95 eV, 4.73 eV, and 8.54 eV are required to remove the first carbon atom from the C(110), C(001), and C(111) clean surfaces, respectively. The high extreme pulling energy of C(111) is consistent with previous experimental and theoretical studies indicating the surface is very stable [12]. On the other hand, the lower stability of the C(110), which has higher surface energy than the other two surfaces, explains the reason for the lower energy cost for the wear processes. Another interesting observation common to all the considered surfaces is that the removal of carbon atoms is likely in the form a carbon chain.

The adsorption of oxygen in ether configuration is unlikely to cause wear of the diamond surfaces through the detachment of the oxygen atoms, which occurs without causing any deformation of the underlying carbon-carbon bonds (**Figure 1b**). The energy trend is similar for all three surfaces, in which the system energy increases as the oxygen atom is pulled up and becomes flat when the oxygen atom is detached from the surface. On the contrary, the pulling up of oxygen atoms in the carbonyl (ketone) group can lead to the detachment of carbon atoms from the surfaces. We found that the wear mechanism depends on the surface orientation. For the C(110) and C(111) surfaces that terminate with zig-zag chains, the displacement of oxygen atoms can cause progressive detachment of the surface chains, which is consistent with experimental findings on wear regime of diamond [25], [26]. Both the force (**Figure S2**) and the energy (**Figure 1c**) are much higher for the C(111) surface, consistently with previous calculations indicating that the C(111) surface is very stable [12]. Meanwhile, the wear on C(001) surface will likely form a CO molecule rather than a carbon chain (**Figure 1c**). Compared to the case of the clean surface, the presence of oxygen atom in the ketone configuration helps to reduce the energy costs for carbon detachment. In particular, the pulling energies required to detach the first carbon atom from the surface are reduced by 0.98 eV, 2.22 eV, and 5.14 eV to values of 1.97 eV, 2.51 eV, and 3.13 eV for the C(110), C(001), and C(111) surfaces. Among the three surface directions, the C(110) is the most easily wearable since the associated energy cost and restoring forces (**Figure S2**) are lower than for the other surfaces.

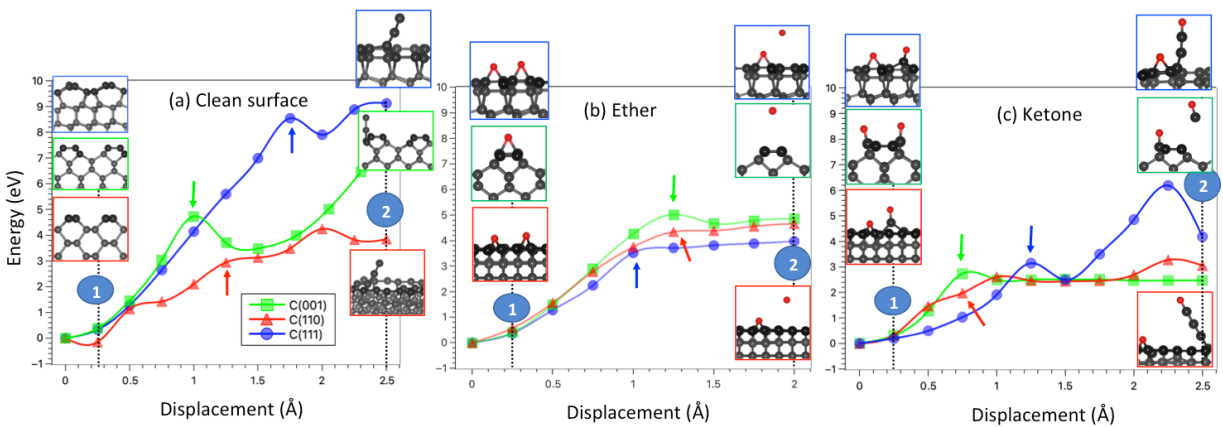

Figure 1. Wear of clean (a), ether (b), and ketone (c) containing diamond surfaces after

A closer look into the reaction path of the wear process can be revealed in **Figure S3** which provides more insight into the atomic mechanism of the most easily wearable C(110) surface. When the oxygen atom is displaced from the surface (step 1 →2), the two C-C bonds are stressed causing an increase in the energy and restoring force acting on the displaced atom. After a C-C bond is broken, the force drops to a lower value. If the O atom displacement is continued, the restoring force on the O atom increases again (step 3→4) until the second C-C bond with the surface is broken. As this mechanism is repeated, the carbon chain becomes longer and longer. It is also worth mentioning that the longer the carbon chain, the longer displacement it takes for the force to increase. The result is due to the elastic properties of the carbon chain that, as a collection of springs added in series, becomes less stiff as a new C-C bond is incorporated in the chain. This phenomenon is also represented in the total energy of the system. At the beginning of the process, the system energy increases at a high rate. However, when the carbon chain becomes longer, the increasing rate of energy becomes smaller indicating that the elastic energy slowly increases since the global spring constant of the chain becomes smaller as the chain length increases. The formation of carbon chain along the zig-zag chain is consistent with previous DFT calculations on C(110) surfaces, showing that the C-C bonds connecting the carbon layer are the weaker and more likely to be broken ones [19].

We finally examine the effects of the presence of other O atoms adsorbed in the surrounding of the ketone group (**Figure 2a**). We found that when an additional O adatom (O3- red) is adsorbed far from the ketone group, the wear mechanism and the associated energy increase is similar to what observed before, i.e., a carbon chain is formed upon the displacement of the O atom forming a double bond with the firstly detached carbon. On the contrary, when an additional O adatom (O4 - blue) forms a bond with a first neighboring site of the ketone group, the detachment of the carbon chain is inhibited and only the C atom belonging to the ketone group is detached through the formation of CO molecule, as observed in the case of the C(001) surface. The result suggested the wear mechanism as well as the length of the C chain is affected by the adsorption of other species around the ketone group.

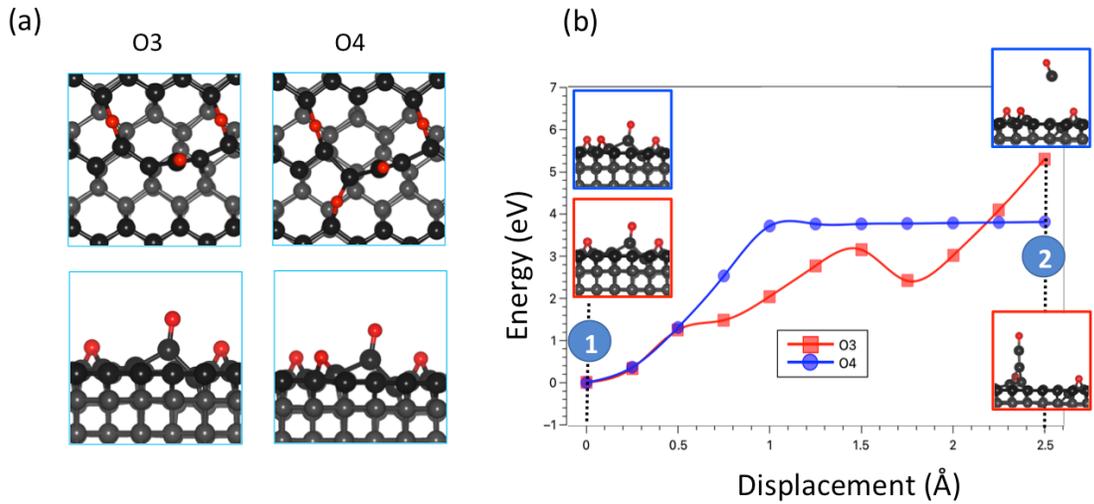

Figure 2. Top and lateral views of the O adatom arrangements considered for the C(110) surface (a). Energy change during the out-of-plane displacement of the O atom belonging to the ketone group (b).

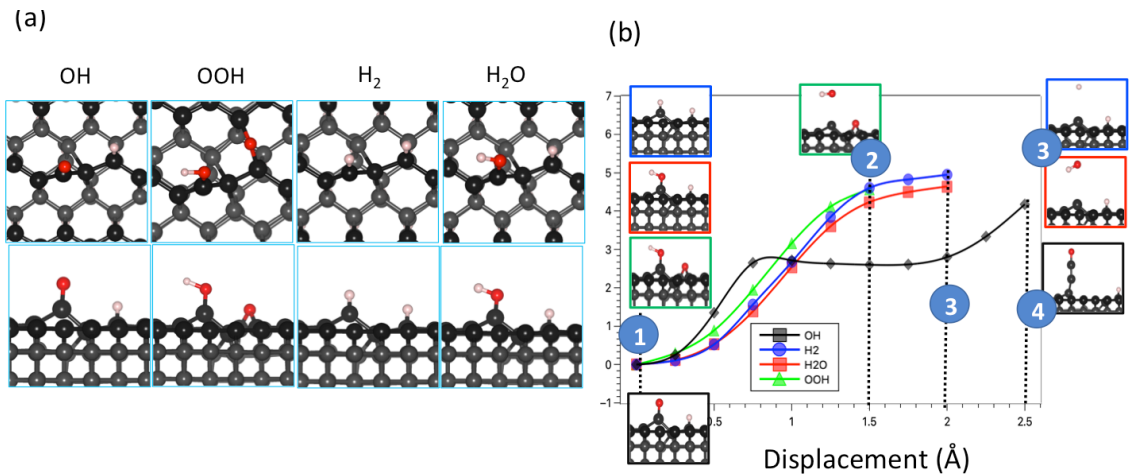

Figure 3. Top and lateral views of the chemical groups adsorbed on the C(110) surface (a). Energy during the out-of-plane displacement of the H, OH and O groups (b).

As our calculation shows that C(110) is the easiest surface orientation to wear, we further extend the study of the wear mechanism on this surface by considering the presence of different chemical species and dissociated molecules, as shown in **Figure 3a**. These species are highly present in the working environment as a result of atmospheric gas dissociative adsorption [11, 37, 38]. We found that when the oxygen atom in the ketone group is replaced by a H atom or OH group, wear cannot be initiated by the out-of-plane displacement of these groups: the displaced atoms are detached from the surface without causing any deformation for the C-C bonds of the substrate. The energy (**Figure 3b**) and force (**Figure S4b**) trends are similar for H/OH. This could be due to the fact that these groups form O-C/H-C single bonds with the substrate. These single bonds are weaker than the O=C double bond presented in the ketone configuration, so the C atom dragged by the pulled atom cannot withstand the restoring force from the substrate. Secondly, if the O forms a double bond with the C atom, it could reduce the strength of other C-C bonds as more electrons are shared to form the O=C double bond. As shown in **Figure 3b** (black line), the energy cost necessary to pulling out of the surface the carbon chain terminating with an O atom (resulting, e.g., from the dissociation of a OH group) is much lower than the energy cost to detach the OH or H groups from the surface.

Overall, the comparison of the behavior of O=C, H-C, OH-C groups, suggests that in order to initiate wear, i.e., produce the detachment of C atoms from the surface, the dragging specie should form a double bond with the displaced carbon atom, as stronger than the restoring C-C bonds.

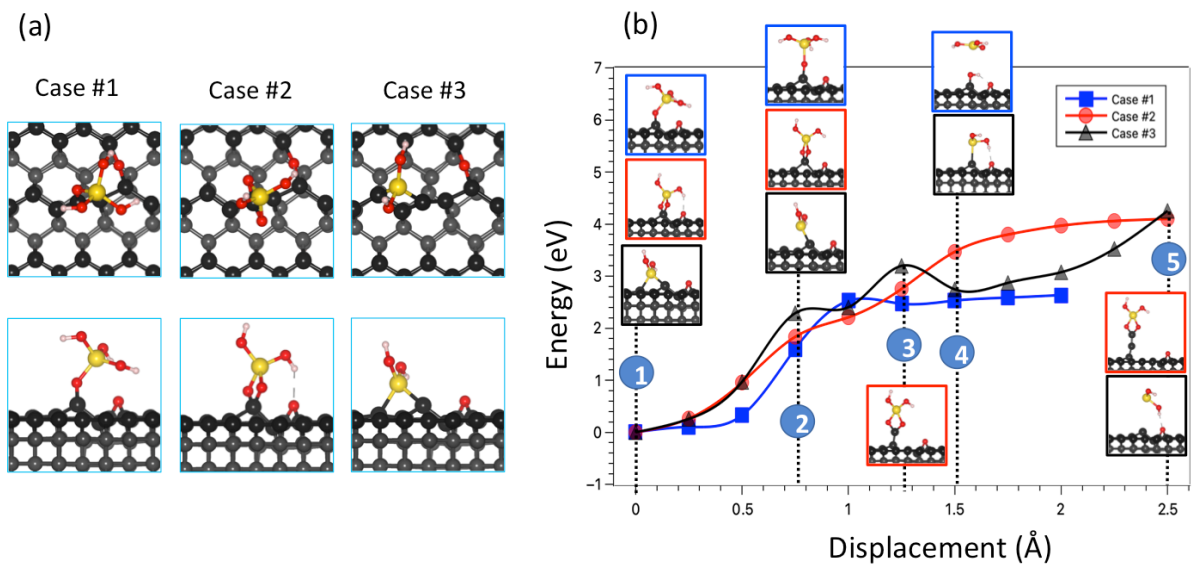

Figure 4. Different initial adsorption configurations of a silicate cluster on the C(110) surface (a). Wear mechanism caused by the detachment of the silicate cluster on C(110) surface (b).

Experimental studies indicate that the sliding of silicate against diamond can wear the latter [7, 9, 39]. To investigate this situation, we adopt a simplified model containing a silicate cluster interacting with the C(110) surface, which was found to be the most easily wearable. We consider three different adsorption configurations of the silicate cluster as shown in **Figure 4a.** In all the initial configurations, the silicate cluster is positioned on top of an out-of-plan carbon atom. In case #1 and #2, one and two single Si-O-C bond(s) is (are) formed with the carbon atom, respectively. In case #3 two Si-C bonds are established with two different C atoms. As shown in **Figure 4b** (case #1 in blue), when the silicate cluster bonds to the surface through a single Si-O-C bond, the Si-O bond is broken after 1.5 Å Si displacement from the original position, leading to a sudden drop of force (**Figure S5**) after which the energy remains constant. When the silicate cluster binds to the C(110) by establishing Si-C bonds with different carbon atoms, **Figure 4b** (case #3 in black), the displacement will cause the breaking of those bonds one by one. As a result, the force (**Figure S5**) on the Si atom shows two peaks as each Si-C bond is maximally stressed. Finally, the silicate cluster is detached at 2.5 A of displacement from the initial position. It is worth mentioning that the detached cluster still interacts with the surface through a hydrogen bond between the OH and the oxygen on the surface causing the energy to increase slowly when the cluster is continuously pulled upward.

On the other hand, when the silicate cluster is bonded with the surface through two Si-O-C single bonds involving the same C atom, known as bidentate structure, the pulling up of the silicate fragment can lead to the formation of a C chain as shown in **Figure 4b** (case #2 in red). This mechanism is similar to that observed when the O atom of the ketone O=C group is pulled, as described above, and we can observe a similar trend in the energy (**Figure 4b**) and force (**Figure S5**) curves. Particularly, the force on the Si atom peaks each time the carbon chain is stressed and drops suddenly when the C-C underlying bond is broken, and the longer the carbon chain, the longer displacement it takes for the force to rise. The system energy increases strongly at the beginning. However, its increase is reduced when the carbon chain becomes longer. The result suggested that the two single bonds on the same C atom could provide a similar effect compared to the C=O double bond in the case of the carbonyl group. That is the formation of two single O-C bonds with the same C atom could help reduce the strength of other C-C bonds as the carbon atom shares more electrons to form the two O-C-Si bonds, leading to the weakening of other C-C bonds. Secondly, the two O-C-Si bonds could provide enough strength to withstand the force when pulling up and avoid the dissociation of the adsorbed species. Therefore, our calculation suggests that in order to initiate wear the carbon atom on the surface needs to form a double bond or at least two single bonds with the counter surface.

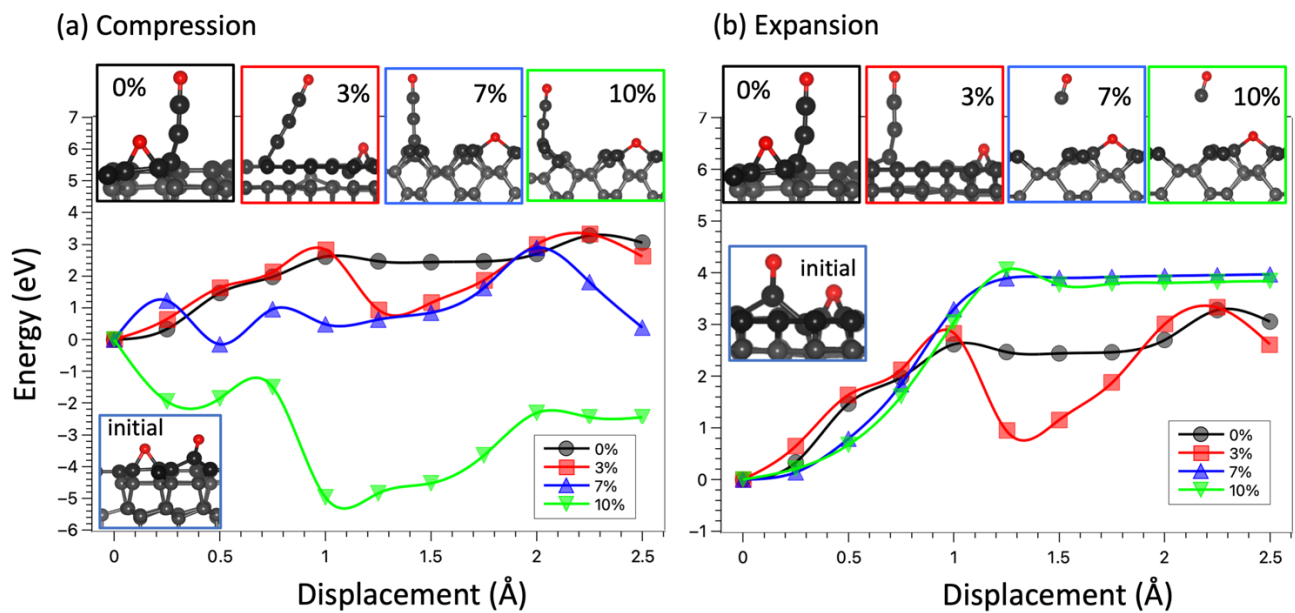

Figure 5. Wear mechanism caused by the adsorption of oxygen on C(110) surface under different stress levels.

In tribological conditions, the surface can be subjected to stress as a result of sliding [25, 26, 40, 41]. Consequently, it is crucial to investigate the impact of stress on the wear mechanism of diamond surfaces. In the present study, stress is introduced by varying the lateral size of the supercell, with four stress levels considered, including 0%, 3%, 7%, and 10%. In the case of compression (as depicted in **Figure 5a,** and the forces in Figure S6), our findings indicate that the wear mechanism remains unchanged, with the detachment of oxygen atoms leading to the formation of long carbon chains. However, there can be significant alterations in energy trends. Specifically, the energy required to detach the oxygen atom decreases as the surface is compressed, particularly under high-stress conditions (7% and 10%). Of particular note, when the surface is compressed at 10%, it becomes energetically favorable for carbon atoms to detach from the surface, as shown in **Figure 5a**. The system energy can be reduced by 4.97 eV in this case.

Therefore, our result suggests that under extremely high load the wear on the diamond surface can become severe.

On the other hand, the wear mechanism can undergo significant alteration when the surface is subjected to expansion. As depicted in Figure 5b, the formation of long C chains remains the predominant wear mechanism at lower stress levels (0% and 3%). However, when the stress level is increased, it fosters the formation of CO molecules, rather than long C chains. Our findings reveal that both energy and force are augmented as the wear mechanism transitions to the formation of CO molecules. Moreover, there is a negligible difference observed between stress levels of 7% and 10%.

## 4. Conclusions

In conclusion, the wear of diamond surfaces is investigated by the mean of first-principles modeling. The comparison between different surface directions, adsorbate, and stress levels in the carbon film has provided novel understanding of the atomistic mechanisms of wear of diamond. Our findings can be summarized as follow.

The primary mechanism of wear involves the detachment of carbon chains from the surface. The energy costs to detach the first C atom depends significantly on the surface orientation, it increases from 2.9 eV for the C(110) and 4.7 eV for the dimer-reconstructed C(001) to 8.5 eV for the Pandey-reconstructed C(111) surface. These results indicate that the C(110) surface, which is the less stable of the low indexes surfaces here considered, is the most easily worn. Meanwhile films exposing the Pandey-reconstructed C(111) surface show the best resistance to wear. The restoring force on the displaced C atom gradually increases until a C-C bond breaks on the diamond surface. This process is repeated, resulting in a longer carbon chain. Interestingly, the elastic properties of the formed chain resemble those of springs added in series: the longer the carbon chain, the lower the restoring force and energy increase.

We modeled the adsorption of O, H and OH and pulled the adsorbed specie to mimic the initiation of a wear process. We observed the formation of a chain just in the case of ketone, C=O, with a lower energy cost than for the clean diamond surfaces. On the contrary, the single adsorbate-C bonds turned out to be less strong than the surface C-C bonds, thus only the pulled adsorbate is detached, without causing any distortion of the underlying carbon surface.

Since diamond can be polished by silica, we investigated the wear mechanisms induced by the adsorption of a silicate. Our results indicate that the strength of single Si-O-C or Si-O bonds is not enough to break C-C bonds in the surface. Two Si-O-C bonds involving the same C atom in a bidentate structure are, instead, able to promote the C detachment from the surface and initiate the wear process through chain formation.

The stress level present in the diamond surface can significantly impact its wear properties. A compressive stress decreases the energy cost for the detachment of carbon atoms, making wear an energetically favorable process at highly stressed conditions. Conversely, when the surface is subjected to expansion, the wear mechanism of carbon chain formation is inhibited and removal of C atoms occurs through the formation of CO molecules at high energy cost.

**SUPORTING INFORMATION**: Computational details, Force changes during the wear of diamond, Detailed wear mechanism on C(110).

**ACKNOWLEDGEMENT**

These results are part of the "Advancing Solid Interface and Lubricants by First Principles Material Design (SLIDE)" project that has received funding from the European Research Council (ERC) under the European Union's Horizon 2020 research and innovation program (Grant agreement No. 865633). PRACE is also acknowledged for computational resources.